\documentclass{article}
\def\beq{\begin{equation}}
\def\eeq{\end{equation}}
\def\bea{\begin{eqnarray}}
\def\eea{\end{eqnarray}}
\def\d{{\mathrm{d}}}

\newfont{\cursive}{pzcmi at 9pt}
\def\~t{\tilde{t}}
\def\Painleve{Painlev\'{e}}

\def\e2phi{\e^{2\Phi}}

\begin{document}

\title{\bf{Black holes without boundaries}}

\author{Alex B. Nielsen
\footnote{\it email: eujin@phya.snu.ac.kr}\\
       {\small\it Center for Theoretical Physics}\\
       {\small\it College of Natural Sciences}\\
       {\small\it Seoul National University}\\
       {\small\it Seoul 151-742, Korea}\\}

\maketitle

\begin{abstract}
We discuss some of the drawbacks of using event horizons to define black holes and suggest ways in which black holes can be described without event horizons, using trapping horizons. We show that these trapping horizons give rise to thermodynamic behavior and possibly Hawking radiation too. This raises the issue of whether the event horizon or the trapping horizon should be seen as the true boundary of a black hole. This difference is important if we believe that quantum gravity will resolve the central singularity of the black hole and clarifies several of the issues associated with black hole thermodynamics and information loss.

\end{abstract}

\section{What is a black hole?}

The laws of black hole thermodynamics have been with us for a long time now. They were first derived in the early 1970's using purely classical general relativity \cite{Bardeen:1973gs}. An important ingredient in the derivation, perhaps the most important ingredient, was the definition of a black hole as the region inside an event horizon. The event horizon, defined mathematically, is the past causal boundary of future null infinity. Even though the concept of a black hole preceded that of an event horizon, because of the work of Hawking and others \cite{HawkingEllis}, the event horizon has come to be synonymous with what we mean by a black hole.

At the time, there were good reasons for defining a black hole this way. Firstly, event horizons capture the idea of a region from which light, and therefore causal signals, can never escape. The event horizon is a smooth continuous surface and it is easy to locate in stationary spacetimes because it is a Killing horizon. In fact, in static, non-rotating spacetimes the event horizon is just the Killing horizon of the time-translational Killing symmetry. As long as one is only interested in the late time behavior of the black hole and the black hole settles down to a stationary state everything works well.

However, event horizons have some troubling features. Due to their global causal teleological nature they are very hard to locate in all but the simplest spacetimes. Numerical relativists have long used marginal surfaces as a proxy for the event horizon since they are much easier to find in numerical simulations. Furthermore, the presence of an event horizon is not physically demonstrable locally, using only local measurements. One first needs to know the entire future history of the spacetime region of interest. This means that the existence or not of event horizons depends, in principle, on the spacetime outside our current past lightcone, something that it is currently impossible for us to know anything about. So long as our past lightcone does not contain a complete Cauchy surface for our entire future we can probably never know if event horizons truly exist!

Apart from these teleological problems, a number of authors have noticed that the laws of black hole thermodynamics can also be derived for structures other than event horizons \cite{Collins:1992,Hayward:1993wb,Ashtekar:2004cn}. These alternative horizons go by a variety of different names but they are all local in nature, since they only depend on properties in some small region of spacetime. In stationary spacetimes, local horizons coincide with event horizons. However, in dynamical spacetimes they do not. This is perhaps most clearly illustrated by the Vaidya spacetime \cite{Poisson:book}. To follow the full evolution of a black hole, one must consider dynamical processes, and here the difference is important.

It is useful to realize that the idea of defining a black hole via the event horizon came before the discovery of Hawking radiation and black hole evaporation. As long as the null energy condition is satisfied and spacetime is regular predictable (without any naked singularities), then the apparent horizon must always lie inside the event horizon \cite{HawkingEllis}. However, Hawking radiation is expected to violate the null energy condition and hence there is no reason to expect the apparent horizon will lie inside the event horizon, especially when the horizon is evaporating. 

This then raises a question. If the laws of black hole thermodynamics refer to something physical and these laws can be derived for two different types of horizons, that on a given hypersurface will typically have different areas and different surface gravities, which one refers to the true thermodynamic behavior? This question has important implications for how we understand black hole evolution and the types of processes that are likely to be relevant.

\section{Thermodynamics with trapping horizons}

Hayward's trapping horizon \cite{Hayward:1993wb} is a conceptually simple version of a locally defined black hole horizon. The trapping horizon is a three-dimensional surface in spacetime whose two dimensional `time-slicings' have the following properties
\begin{enumerate}
\item[\it{i}.] The ingoing null rays are converging
\item[\it{ii}.] The outgoing null rays are converging inside the horizon and diverging outside the horizon
\end{enumerate}\bigskip
These conditions are satisfied by the event horizon in the Kerr-Newman class of solutions. More generally, any spherically symmetric (not necessarily static) metric in four dimensions can be put in the form \cite{Nielsen:2005af}
\bea \d s^2 & = & - e^{-2\tilde{\Phi}(\tau,r)}\left(1-\frac{2m(\tau,r)}{r}\right)\d \tau^{2} + \nonumber \\ & & + 2e^{-\tilde{\Phi}(\tau,r)}\sqrt{\frac{2m(\tau,r )}{r}}+\d r^{2} + r^{2}\d\Omega^{2} \eea
in so-called \Painleve-Gullstrand coordinates, where $m(\tau,r)$ is immediately recognizable as the Misner-Sharp mass function. The Misner-Sharp mass function can be interpreted as the quasi-local mass contained within a sphere or radius $r$ at a time $\tau$. The radial null rays $l$ and $n$ have the following expansions
\beq \theta_{l} = \frac{2}{r}\left(1-\sqrt{\frac{2m(\tau,r)}{r}}\right), \eeq
\beq \theta_{n} = -\frac{1}{r}\left(1+\sqrt{\frac{2m(\tau,r)}{r}}\right). \eeq
Thus we see that the the surface $r=2m(\tau,r)$ is a trapping horizon\footnote{Modulo a few caveats \cite{Nielsen:2005af}.}. Note that in general, the surface $r=2m(\tau,r)$ is {\it{not}} the location of the event horizon. At the horizon we have
\beq \label{firstlaw} \frac{\partial m}{\partial \tau} = \frac{1}{8\pi}\frac{(1-2m')}{2r}\frac{\d A}{\d \tau}, \eeq
where $m'=\frac{\partial m}{\partial r}$ and $\tau$ is a parameter labeling `time-slicings' of the horizon. This equation has the same form as the first law of black hole thermodynamics $\d m = \frac{1}{8\pi}\kappa\;\d A$ with a surface gravity that agrees with other definitions of surface gravity \cite{Nielsen:2007ac} and a classically defined temperature (assuming the entropy is $A/4$) of
\beq \label{temperatureclassical} T = \frac{1}{2\pi}\frac{(1-2m'_{H})}{2r_{H}}. \eeq
In order to obtain a version of the second law we can just compute $G_{ab}l^{a}l^{b}$, where $G_{ab}$ is the Einstein tensor. This gives
\beq G_{ab}l^{a}l^{b} = \frac{2e^{\Phi}}{r^{2}}\frac{\partial m}{\partial \tau}\sqrt{\frac{2m}{r}} - \frac{2}{r}\frac{\partial\Phi}{\partial r}\left(1-\sqrt{\frac{2m}{r}}\right)^{2}. \eeq
Rearranging, and imposing (\ref{firstlaw}) at $r=2m(\tau,r)$ we find
\beq \frac{\partial A}{\partial \tau} = \frac{16\pi r^{3}e^{-\Phi}}{1-2m'} G_{ab}l^{a}l^{b}. \eeq
Thus we see that the area of the horizon $A$ is increasing if $G_{ab}l^{a}l^{b} > 0$. By the Einstein equations we can write this condition as $T_{ab}l^{a}l^{b} > 0$, which is just the null energy condition (NEC). The area of the horizon is increasing if the NEC is satisfied and can decrease only if the NEC is violated.

So much for the analogous classical laws of black hole mechanics. What really clinched the idea of black hole thermodynamics was the discovery of the quantum Hawking effect. Could trapping horizons also generate Hawking radiation? A derivation of the Hawking effect given by Parikh and Wilczek \cite{Parikh:1999mf,Di Criscienzo:2007fm} demonstrates that this is indeed the case. To lowest order, the Klein-Gordon equation for a massless scalar field $\phi$, of the form
\beq \phi = e^{-iS(\tau, r)/\hbar}, \eeq
reduces to the Hamilton-Jacobi equation,
\beq g^{ab}\partial_{a}S\partial_{b}S = 0. \eeq
For this scalar field one can calculate a tunneling rate given by
\beq \Gamma \sim \phi\phi^{*} = e^{-2\mathrm{Im}\; S}. \eeq
The imaginary part of $S$ turns out to come from a simple pole at $r=2m(\tau,r)$. From the geometrical optics approximation we have
\beq \label{geooptics} S(\tau,r) = \omega \tau - \int k(r)\d r, \eeq
and hence
\beq S \sim \omega \tau + \frac{2r_{H}\omega e^{\Phi_{H}}}{\left(1-2m'_{H}\right)}\int\frac{\d r}{\left(r-r_{H}\right)}. \eeq
This integral can be performed for outgoing modes moving from inside the horizon to outside by deforming the contour into the lower half of the complex plane, which gives a complex contribution to $S$
\beq \textrm{Im}S = \frac{4\pi r_{H}\omega e^{\Phi_{H}}}{\left(1-2m'_{H}\right)}. \eeq
Thus we get a thermal tunneling rate with a temperature given by
\beq \label{temperaturequantum} T = \frac{1}{2\pi}\frac{e^{-\Phi_{H}}}{2r_{H}}\left(1-2m'_{H}\right). \eeq
This calculation suggests that the trapping horizon at $r=2m(\tau,r)$ can generate a thermal flux of Hawking radiation even when there is no event horizon present \footnote{A different derivation of the Hawking effect, but equally independent of the global structure, was given in \cite{Peltola:2008jx}.}. This derivation should be valid in the semi-classical regime as long as the geometrical optics approximation (\ref{geooptics}) is valid. Typically this will be valid till much later in the black hole's evaporation than the stationarity assumption \cite{Visser:2001kq}.

There are two interesting things to note about this procedure. Firstly, note that the `quantum' temperature (\ref{temperaturequantum}) is not quite the same as the `classical' temperature (\ref{temperatureclassical}) as they differ by a factor of $e^{-\Phi_{H}}$. In many simple situations, such as Schwarzschild and Reissner-Nordstr\"{o}m, this factor is just unity. However, in other models it does make a difference, for example \cite{Nielsen:2006gb}. This factor is not just a gauge choice and cannot be removed by a redefinition of the time coordinate since the time coordinate is fixed at infinity to be the proper time of static observers. Some speculation on a possible resolution of this difference is given in \cite{Hayward:2008jq}.

The other subtle point is that both (\ref{temperaturequantum}) and (\ref{temperatureclassical}) depend on the choice of foliation since the partial derivative $m'$ is taken in the direction of constant $\tau$. This dependence arises in the first law because one is using $\tau$ to foliate the horizon and in the quantum case because one is using $\tau$ to pick out a notion of energy. This dependence can most likely be traced back to the requirement in a quantum theory to choose foliations in order to impose equal time commutation relations.

\section{Black hole evaporation}

The importance of the trapping horizon for the black hole hole information paradox is clear. In the original conception, the collapse of a star to a black hole led to the formation of a singularity and a boundary of spacetime.  If there is a boundary then the world line of particles will come to an end at finite parameter time (even if the boundary is not singular). The pair creation of virtual particles at the horizon leads to Hawking radiation when one of the virtual particles falls into the black hole with negative energy, allowing the other particle to carry positive energy away to infinity. It is this process that leads to pure states turning into mixed states, since the outgoing Hawking radiation particles are entangled with the infalling particles until the infalling particles cease to exist \cite{Mathur:2008wi}. The entangled particles just `fall off' the edge of spacetime and the Hawking radiation particles are entangled with nothing, becoming a fundamentally mixed state.

This boundary of spacetime is closely related to the event horizon. If the worldines of particles don't end after finite parameter time, then they can only `end' after infinite parameter time, at infinity, or cease to be well defined in some quantum geometry. If all worldlines reach infinity then there is no event horizon. Spacetimes with both spacelike boundaries where the worldlines of particles come to an end and asymptotically flat regions where they don't, will have event horizons. If there is a spacelike boundary then entangled particles will `fall off', leading to the non-unitary evolution of states. 

If we don't feel bound to define black holes by event horizons, but use trapping horizons instead, then it is perfectly reasonable to have a black hole without a spacelike boundary. If there is information inside the event horizon that we want to get out, then by definition, the only way to get it out is by some non-local process. This is not the case with trapping horizons. Trapping horizons are not causal boundaries and can be crossed from the inside to the outside when they are evaporating and timelike.

Of course, we cannot follow the evolution using the formalism above once the semi-classical approximation breaks down. Quantum gravity must come in at some point and tell us whether a spacelike boundary really does form and by extension an event horizon. But whatever the ultimate resolution of the black hole information paradox is, be it in the context of string theory \cite{Mathur:2008wi}, loop quantum gravity \cite{Ashtekar:2005qt} or something else, trapping horizons provide a classical framework for dealing with black holes without boundaries and event horizons \cite{Nielsen:2008kd}. 

\section{Clarifying black holes}

The local black hole horizons program has already led to a number of insights into black holes. The relativity community has been partly conscious of some of these points for a while now, while others have received much greater emphasis only recently.
\begin{enumerate}
\item[\it{i}.] Astrophysical black holes do not require event horizons. The astrophysical data can be explained without technical event horizons since the experiments can only probe local gravitational effects. In fact, astrophysical experiments cannot, by their very nature, demonstrate the existence of a true event horizon.
\item[\it{ii}.] The black hole area increase law does not require an event horizon. In fact, the area increase law can be proven for several different types of horizon (event horizons included) and may not be as fundamental as once thought.
\item[\it{iii}.] Hawking radiation does not require an event horizon \cite{Visser:2001kq}. This should sound reasonable as the Hawking effect is only a result of local quantum field theory on a local curved manifold. In fact, the Hawking effect does not even require the Einstein equations to hold. 
\item[\it{iv}.] The classical surface gravity, derived from a dynamical law for energy flux across a horizon, does not always correspond to the temperature derived from an explicit calculation of the Hawking effect\cite{Nielsen:2007ac}. This is true even in static cases once one accepts that the classical surface gravity should be defined locally.
\item[\it{v}.] Information loss, as first conceived by Hawking \cite{Hawking:1976} and expressed in \cite{Mathur:2008wi}, does require an event horizon. The tracing over unknowable states, that leads a pure state to evolve to a mixed state, is explicitly tied up with the existence of an event horizon.
\item[\it{vi}.] Locally defined black hole horizons can be timelike. This means that it is perfectly possible for causal signals to travel from inside the black hole to outside, once the black hole has begun to evaporate. 
\end{enumerate}\bigskip

\section{Conclusions}

We have seen that there are many tantalising reasons to take the the idea of black holes defined in terms of trapping horizons seriously. This ranges from their astrophysical behaviour, to their thermodynamic properties and the production of Hawking radiation. In addition they offer several advantages over event horizons in that they can be located locally, cannot occur in flat space and do not seem to be subject to the same stringent requirements for information loss.

However, many questions still remain unanswered. Technical questions about the uniqueness and essential properties of trapping horizons have only be lightly researched. The possibility of forming trapping horizons in a spacetime without an event horizon or a central singularity have only been studied in very heuristic models without a rigorous investigation of the effect of energy condition violating Hawking radiation \cite{Roman:1983zza,Hayward:2005gi}. The possibility of traveling over a timelike trapping horizon from the inside to the outside should also be investigated more clearly. When is it possible and for how long?

The issue of thermodynamics applied to trapping horizons has been researched in more detail. However, there remain some open issues. Firstly, the question of the surface gravity and its apparent mismatch with the temperature of the Hawking radiation \cite{Nielsen:2007ac}. Secondly, there is the possibility that the meeting of an inner and horizon trapping horizon might result in a violation of the third law, as the surface gravity goes instantaneously to zero \cite{Hayward:2005gi}. Thirdly, and perhaps most importantly, one can ask about the assignment of entropy to trapping horizons. If outer trapping horizons with non-zero area can be suddenly created in dynamical collapse scenarios (and there is good evidence that this is possible \cite{Booth:2005ng}) then should one assign a sudden discontinuous increase in the generalised entropy in this case? Perhaps more worryingly, if outer trapping horizons can disappear suddenly with a finite area, how should one account for the apparent sudden discontinuous drop in the entropy, unless this process is accompanied by a sudden discontinuous burst of Hawking radiation?

Despite these unanswered questions there are good grounds to believe that the trapping horizon paradigm has much to teach us about black holes. Given the above listed points, it seems that unitary evolution of black holes is perfectly possible without needing to resort to large-scale violation of locality or duplication of information. When one searches for clues to quantum gravity from black hole behaviour, one must be careful what questions one asks. As Sherlock Holmes would no doubt agree, when concluding improbable outcomes one must be very careful that one has eliminated as impossible all other possibilities.

\section*{Acknowledgements}

This work was supported by the Korea Research Foundation Grants funded by the Korean Government (MOEHRD) KRF-2007-314-C00055 and KRF-2008-314-C00069.

\end{document}